\documentclass[12pt,a4paper]{article}
\usepackage{a4wide}
\usepackage{latexsym}
\usepackage{epsf}
\usepackage{amssymb}
\begin{document}
\def\be{\begin{equation}}
\def\ee{\end{equation}}
\def\bea{\begin{eqnarray}}
\def\eea{\end{eqnarray}}

\def\pd{\partial}
\def\a{\alpha}
\def\b{\beta}
\def\g{\gamma}
\def\d{\delta}
\def\m{\mu}
\def\n{\nu}
\def\t{\tau}
\def\l{\lambda}

\def\s{\sigma}
\def\e{\epsilon}
\def\scri{\mathcal{J}}
\def\cM{\mathcal{M}}
\def\tcM{\tilde{\mathcal{M}}}
\def\RR{\mathbb{R}}

\hyphenation{re-pa-ra-me-tri-za-tion}
\hyphenation{trans-for-ma-tions}


\begin{flushright}
IFT-UAM/CSIC-98-20\\
hep-th/9807226\\
\end{flushright}

\vspace{1cm}

\begin{center}

{\bf\Large Geometric Holography, the Renormalization Group and the c-Theorem}

\vspace{.5cm}

{\bf Enrique \'Alvarez}${}^{\diamondsuit,\clubsuit}$
\footnote{E-mail: {\tt enrial@daniel.ft.uam.es}}
{\bf and C\'esar G\'omez}${}^{\diamondsuit,\spadesuit}$
\footnote{E-mail: {\tt iffgomez@roca.csic.es}} \\
\vspace{.3cm}

\vskip 0.4cm

${}^{\diamondsuit}$\ {\it CERN-TH, 1211 Geneva 23, Switzerland, and Instituto de F\'{\i}sica Te\'orica, C-XVI,
  Universidad Aut\'onoma de Madrid \\
  E-28049-Madrid, Spain}\footnote{Unidad de Investigaci\'on Asociada
  al Centro de F\'{\i}sica Miguel Catal\'an (C.S.I.C.)}

\vskip 0.2cm

${}^{\clubsuit}$\ {\it Departamento de F\'{\i}sica Te\'orica, C-XI,
  Universidad Aut\'onoma de Madrid \\
  E-28049-Madrid, Spain}

\vskip 0.2cm

${}^{\spadesuit}$\ {\it I.M.A.F.F., C.S.I.C., Calle de Serrano 113\\ 
E-28006-Madrid, Spain}

\vskip 1cm


{\bf Abstract}

\end{center}

\begin{quote}
  
  In this paper the whole geometrical set-up giving a conformally invariant holographic projection
of a diffeomorphism invariant bulk theory is clarified. By studying the renormalization group flow 
along null geodesic congruences
a holographic version of  Zamolodchikov's c-theorem is proven.
  
\end{quote}


\newpage

\setcounter{page}{1}
\setcounter{footnote}{1}
\renewcommand{\theequation}{\thesection.\arabic{equation}}


\section*{Introduction}

Holography, as originally suggested by 't Hooft \cite{'th} postulates that {\em given a closed surface, we can represent
 all that happens inside it by degrees of freedom on this surface itself}. This postulate implies that quantum
 gravity is an special type of quantum field theory in which all physical degrees of freedom
can be holographically projected onto the boundary. The fundamental aim
of holography was to reconcile gravitational collapse and unitarity of quantum mechanics at the 
Planckian scale. 
\par
Taking string theory ( or M-theory ) as the most natural candidate for a  quantum theory of gravity, holography suggests
looking  for a new description in terms of holographic degrees of freedom living at the boundary
of space time. There are until now two independent attemps in this direction, namely
the M(atrix) formulation of M theory and the Anti de Sitter/Conformal Field Theory (AdS/CFT)  pairs based on
Maldacena's conjecture \cite{malda}. It is in this second case that a concrete form of holographic projection can
be rigorously stablished. In the  simplest AdS/CFT framework (\cite{witten}) the theories that are holographically related
are Type IIB string theory on AdS and N=4 pure Super Yang Mills theory at the boundary.
\par
The discovery of these ( gravity/gauge theory) holographic pairs opens the door to a different use of
 holographic ideas, namely in the direction of finding the gravitational description of pure
gauge dynamics and reciprocally. In this approach the first problem
to be addressed is that of finding, given a gauge theory on a spacetime manifold M, a suitable holographic
 manifold $\tilde{M}$  with boundary $M$. Presented the holographic problem in this way the appareance of CFT's at the
boundary is not accidental but very much dependent on the underlying geometry. In fact
, as pointed out many years ago by Penrose, the very definition of (conformal)  spacetime boundary is done
in general only up to conformal transformations. More precisely the bulk geometry only fixes a conformal
 class of geometries at the boundary, forcing us to use theories at the boundary invariant under
 conformal transformations.
\par
 Thus the
first step in the construction of holographic pairs, consists in fixing a bulk geometry in terms of
 a given conformal class at the boundary. This mathematical problem has been worked out in
reference \cite{fg},  leading to the conclusion that bulk geometry is uniquely determined by conformal data
 at the boundary if the bulk space solves vacuum Einstein's  equations with non vanishing and negative
 cosmological
constant i.e AdS space time. From the physical point of view
a necessary condition to achieve uniqueness is to impose vanishing gravitational Bondi news through
the boundary. This requirement implies in general the existence of confining gravitational fields
 preventing all massive degrees of freedom to reach the boundary. This is a very important dynamical
 ingredient in the construction of holographic pairs where some kind of
Kaluza Klein dimensional reduction is involved. In fact
decoupling of Kaluza Klein modes is very much dependent on the fact that these modes are locked by
 the confining
gravitational field \cite{br}.

\par
In all approaches to holography a physical question arises naturally, namely the meaning of the
 extra holographic space time dimension. Based on the original Maldacena's description of near horizon
 geometry for D-branes, it seems natural to identify this extra holographic dimension as parametrizing
 somehow  the renormalization group scale of the theory. From this perspective the Einstein equations
 governing the bulk geometry, are promoted to renormalization group equations; a situation that strongly
resembles what take place in string theory. In order to get some progress in this direction we must
 certainly go a trifle beyond the AdS/CFT type of holographic pairs.
\par
Our approach to the problem will be as follows. We will consider as holographic bulk manifold
Einstein spaces which are asymptotically AdS and with no
Bondi news at the conformal infinity.
\par
 Secondly we will use as holographic equations describing the
variation of the metric in terms of the holographic coordinate- the transport of a $(d-2)$-dimensional surface
along null geodesic congruences  ending at the conformal infinity.
\par
 Finally we will interpret
 the area of these surfaces, in an appropiated normalization, as the holographic definition of
 Zamolodchikov's c-function
with the affine parameter -along the null congruence entering into the bulk- as renormalization
 group variable.
\par
 This definition of the c- function is based on the IR/UV interpretation of the AdS/CFT holographic pairs \cite{sw}, and coincides,
 in the conformal case, with the central extension. In this set up
we can relate the predicted (by the c-theorem) behaviour of the c-function to very general gravitational properties
of null congruences. More precisely the c-theorem appears , in this context, as a quite direct consequence of Raychadhuri's 
 equation. The physical meaning of this result
goes into the direction of translating general properties of quantum
field theories into gravitational language. 
\par
An optimistic interpretation of the results
presented in this paper would  indicate a deep interplay between the irreversibility of the renormalization
group flow and the famous singularity theorems in general relativity \cite{he}, where now the
 singularity theorems are worked out for the holographic bulk space.
\par
Next let us summarize the outline of the paper. In section 1 we review the geometrical
set up of holography, stressing the two different approaches based on Lorentz or Poincar\'e metrics. We also 
review some generalities of Penrose's construction of conformal infinity and discuss
a possible extension of geometric holography to non local loop variables. In section 2 we address the
question of Bondi news in terms of vanishing Bach tensor at the boundary. In sections 3 and 4 we discuss
the holographic definition of Zamolodchikov c-function and finally in section 5 we give a purely gravitational
proof of the c-theorem. We end the paper with some general comments and speculations.


\section{Geometric Holography}

The aim of this section is to explore the geometric setting of the holographic Anti de Sitter/Conformal Field Theory
 (AdS/CFT)  relationship (\cite{gkp}\cite{malda}\cite{witten}). Heavy use will be made for this 
purpose of the mathematical results contained in the paper of  Fefferman and Graham \cite{fg}.
\par
In strict mathematical terms the problem associated with the holographic projection
is that of finding the conformal invariants of a given manifold $M$, with generic
signature $(p,q)\equiv((-1)^p,1^q)$ and dimension $n = p + q$, in terms of the Riemannian invariants 
of some other manifold $\tilde{M}$
in which $M$ is  contained in some precise sense.
In order to employ a more physical terminology, we will refer to $M$ as the {\em space-time}
and to $\tilde{M}$ as the {\em bulk} or {\em ambient space}.
\par
Following \cite{fg} we will work out this geometrical problem from two different points of view, depending
on the dimension and signature of the bulk space. In the so-called {\em Lorentzian } approach the 
ambient space will have signature
$(p+1,q+1)$ while in the second approach, based on Penrose's definition of conformal infinity the bulk space will have
either $(p,q+1)$ or $(p+1,q)$ signature, leading to two different kinematical types of geometric holography.

\subsection{Lorentz Holography}

Given a Riemannian manifold $M$, a conformal class $[g]$ on $M$ is defined as the equivalence 
class of metrics: $g_{\m\n}\sim g'_{\m\n}$ iff $g'_{\m\n} = \Omega^2 g_{\m\n}$ , 
$\Omega$ being a smooth  function on $M$.
\par
Conformal invariant tensors $P(g)$ are defined by the transformation law;
\be
P(\lambda g) = \lambda^{-\Delta} P(g)
\ee
where $\Delta$ is the conformal weight; they are thus associated with a given conformal class,
$[g]$. 
\par
In order to describe a conformal class $[g]$ on $M$, let us introduce the space $G$ 
\footnote{A ray sub-bundle $G$ of the bundle of symmetric 2-tensors on $M$.}
defined in terms of some reference metric $ds^2_{ref}$ as:
\be
G \equiv \{ (x, t^2 g^{(ref)}_{\m\n}(x)); t>0; x\in M \}
\ee
In fact any {\em section} of $G$ (i.e., any map from $M$ to $G$ specified by
a function $t(x)$) defines a particular representative in the conformal class $[g]$.
Conversely, any metric $g_{\m\n}\in [g]$ on $M$ defines an imbedding of $M$ into $G$.
Any conformal transformation on the metric can in this way be interpreted
as moving from one section to another. To be specific, we define a {\em dilatation} as:
\be
\delta_s (x,\bar{g}_{\m\n}) \rightarrow (x,s^2 \bar{g}_{\m\n});(x,\bar{g}_{\m\n})\in G
\ee
The big space $G$ is naturally endowed with a natural (degenerate) metric, here denoted by $g_0$,
inherited from the one
defined at $M$ \footnote{In mathematical terms, this is equivalent to:
 $g_0 (\vec{x},\vec{y}) = \bar{g}(\pi_{*}\vec{x},\pi_{*}\vec{y})$
for $\vec{x},\vec{y}\in T_{(x,\bar{g})}(G)$, tangent vectors to $G$ at the point $(x,\bar{g})$,
and $\pi_{*}$ denoting the pull-back of the natural projection to the base, 
$ \pi(x,\bar{g})\equiv x\in M$.}
. In terms of the local coordinates $(x^{\m}, t)$
\be
ds_0^2 (x,t) \equiv  t^2 ds_{(ref)}^{ 2}(x)
\ee

 In this generalized set-up, conformal fields (and conformal weights)
on $M$ will be defined by their transformation law with respect to $\delta_s$.
\par
If we write $\delta_s \equiv e^{s\vec{T}}$ ($\vec{T}$ thereby denoting the corresponding infinitesimal
generator), then it is plain that $\vec{T}$  is a null vector and, in fact, orthogonal to any other tangent vector, $\vec{X}\in T_G$
\be
g_0 (\vec{T},\vec{T}) = g_0 (\vec{T},\vec{X}) =0
\ee
(because it acts trivially on the points of the base, $x\in M$). 
\par

\begin{figure}[!ht] 
\begin{center} 
\leavevmode 
\epsfxsize= 10cm
\epsffile{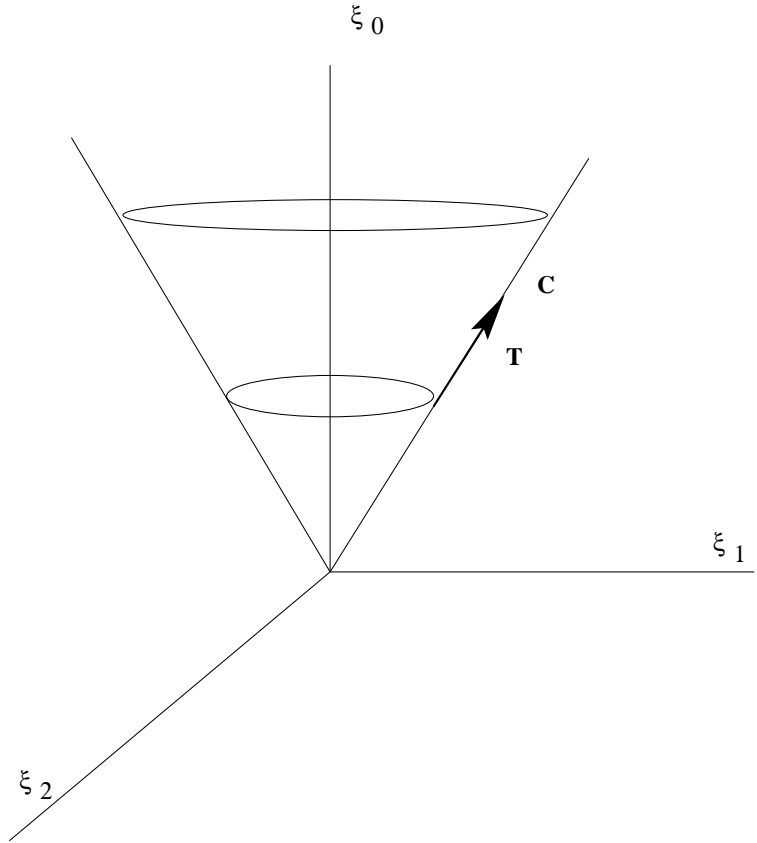} 
\caption{ Geometrical representation of the space G  corresponding to $M = S^1$, as the cone $\mathcal{C}$.}
\label{fig:uno} 
\end{center} 
\end{figure}

In order to get a geometrical image of the space $G$, let us consider a simple example of the previous construction. Let $M$
be the one-dimensional sphere $S^1$ of unit radius, with the metric induced by the imbedding in $\mathbb{R}^3$,  called
 $ ds^2$.
Conformal transformations, $ds^2\rightarrow t^2 ds^2$ are now equivalent to changes in the radius of the 1-sphere.
The big space $G$ defined above is thus in this case a cone $\mathcal{C}$, as shown in the Figure ~\ref{fig:uno} with 
each transversal section being identified with a copy of $S^1$ with a particular metric in the conformal class. The vector
$\vec{T}$ generating dilatations goes along the generatrix of the cone $\mathcal{C}$ connecting the same point in the one-sphere $S^1$
in different transversal sections.

Once we have this picture in mind, let us proceed to define the  ambient space as
 the product of our big space $G$ with a closed segment:
\be
\tilde{G}\equiv G \times [-1,1]
\ee
and we formally identify our previous $G$ with a subset of the new construct\footnote{Which is the common 
boundary of the two complementary regions $\tilde{G}^{+}\equiv G\times [0,1]$
and $\tilde{G}^{-}\equiv G\times [- 1,0]$, which can be visualized in the figure as the interior and exterior regions
to the cone $\mathcal{C}$  Incidentaly,
this defines the natural inclusion $ \iota : G\rightarrow \tilde{G}$.}
: $G\equiv G\times \{0\}$.

It remains to define a natural metric $d\tilde{s}^2$ in the bulk space $\tilde{G}$.
We would like to define it in such a way that any Riemannian invariant $P(\tilde{g})$ on $\tilde{G}$, induce,
when reduced to the original manifold $M$, conformally invariant tensors on $M$.
The way to reduce tensors in the bulk to tensors in the manifold is simplicity itself:  First of all, to reduce
from $\tilde{G}$ to $G$, we use the trivial imbedding, so that in a system of local coordinates
$x^M\equiv(x^{\m},t,\rho)$, with $\rho\in\tilde{G}/G \sim [- 1,1]$, we just put $\rho = 0$
\be
\tilde{P}(\tilde{g})|_M \equiv \tilde{P}(\tilde{g}(\rho = 0))
\ee
Next, to go from $G$ to $M$, we use the imbedding given by the metric $g_{\m\n}$ itself
\be
g_{\m\n}(x)\rightarrow t(x)
\ee
defining in this way a tensor in $M$, which we shall call $\tilde{P}$
\footnote{ In mathematical language, it is just the pull-back under the inclusion:
$\tilde{P}(g)\equiv g^{*}(\tilde{P}(\tilde{g})|_G $. That is:
$\tilde{P}\equiv P(\tilde{g}(x,t(x),\rho=0))$}
Remarkably enough,
Fefferman and Graham were able to prove that $\tilde{P}(g)$ will be conformally invariant
provided that
the following three conditions \footnote{In (more precise) mathematical language, the three conditions above read:
\begin{eqnarray}
i)& \iota^{*} \tilde{g} = g_0 \nonumber\\
ii)& \delta_s^{*} \tilde{g} = s^2 \tilde{g}\nonumber\\
iii)& Ricci(\tilde{g}) = 0
\end{eqnarray}
}
are fulfilled:
\begin{eqnarray}
i)& d\tilde{s}^2(\rho = 0) = ds_0^2 \nonumber\\
ii)& d\tilde{s}^2(x,\lambda t,\rho) = \lambda^2 d\tilde{s}^2(x,t,\rho)\nonumber\\
iii)& R_{MN}(\tilde{g}) = 0
\end{eqnarray}

The first condition simply means that $\tilde{g}$ reduces on $G$ to the metric $g_0$ defined bt the conformal
class $[g]$.
Actually, this first condition $i)$ alone (plus the fact that $\vec{T}^2 = 0$ ) forces the signature of $\tilde{G}$
to be $(p+1,q+1)$ (i.e., of the two {\em extra dimensions} that $\tilde{G}$
gets over those of $M$, one must necessarily be spacelike, and the other one timelike ).
\par
Let us come back to our simple example in order to visualize this result. In this picture the  ambient space is defined in terms
of the (projective) coordinates $\xi^0,\xi^1,\xi^2$ (so that the sphere $S^1\equiv (x^1)^2 + (x^2)^2 = 1$ is mapped onto
$\xi_0^2 -\xi_1^2 - \xi_2^2 = 0$
by means of  $x^i\equiv\frac{\xi^i}{\xi^0}$($i=1,2$)). The dilatation generator on the subspace $G$ (i.e., on the cone 
$\mathcal{C}$)
is forced to be a null vector with respect to the metric $\tilde{g}$ (restricted to $G$). All this is obviously equivalent to
identify the cone $\mathcal{C}$ with the {\em light cone} of the three-dimensional ambient space with signature $(1,2)$
(where the coordinate $\xi_0$ is a {\em time}). Generically, one goes from signature $(p,q)$ in $M$, to signature
$(p+1,q+1)$ for the Lorentzian ambient space.
\par
The second condition ii) above, in turn, is crucial in order to get conformal invariants from Riemannian ones. In fact, this condition
tells how the metric $\tilde{g}$ transforms with respect to those changes of coordinates in $\tilde{G}$ that induce
 in $M$ conformal transformations of the metric. In particular, this implies that Riemannian tensors on $\tilde{G}$ 
are homogeneous in $s$ with respect to dilatations $\delta_s$. This implies that $\tilde{P}(g)$ will be a conformal tensor provided that the
dependence of $\tilde{g}$ on $g_0$ is analytic (that is, if $\tilde{g}(g_0)$ admits a power series expansion).
\par
Precisely in order to control this last condition condition iii) above has been introduced. In fact, the Ricci 
flat condition on $\tilde{G}$
can be thought of as defining a Cauchy problem on the space of metrics with initial condition given by the conformal class $[g]$.
By introducing a formal parameter as above, $\rho\in[- 1,1]$, what we get by solving the Ricci-flatness condition
is a family of metrics $g_{\m\n}(\bar{x},\rho)$, with $\bar{x}$ denoting coordinates in $M$, and such that 
$g(\bar{x},\rho = 0)\in [g]$.
It is the variable $\rho$ the one that could be most properly interpreted as a holographic coordinate, and the preceding 
equations are then describing the {\em holographic flow} of the spacetime metric. Once
 we get a solution of the above problem, we have
a systematic
way of relating quantities defoned on $M$ (the CFT  side) and ambient quantities on $\tilde{G}$
 (the supergravity side), but we shall refrain from entering into this until we have reviewed Penrose's conformal infinity
approach to holography and its precise relationship with the Lorentzian approch discussed in this section.


\subsection{Penrose's Conformal Infinity}
Penrose's construction of conformal infinity (cf. \cite{penrose}) was originally designed 
as a general procedure to address the physics of the asymptotic structure of the space
time in General Relativity. The main idea used to bring infinity back to a finite distance
is to Weyl rescale the physical metric, looking for a new manifold 
$\tilde{\mathcal{M}}$ with boundary, such that the interior of $\tilde{\mathcal{M}}$
coincides with our original spacetime $\mathcal{M}$, endowed with the metric $\hat{g}_{\m\n}$
to be defined in a moment.
In the physics literature it is costumary to denote  the part of $\partial\mathcal{M}$ 
corresponding to end-points of null geodesics by $\mathcal{J}$
\par
Based on the previous construction we will say that a spacetime $(\mathcal{M},g_{\m\n})$ is asymptotically simple
(acually we will be a little bit more general, and allow for asimptotically Einstein behavior)
if there exists a smooth manifold $\tilde{\mathcal{M}}$, with metric $\hat{g}_{\m\n}$, and
a smooth scalar field $\Omega(x)$ defined in $\mathcal{M}$ such that:
\par
i) $\mathcal{M}$ is the interior of $\tilde{\mathcal{M}}$ .
\par
ii)$\hat{g}_{\m\n} = \Omega^2(x) g_{\m\n}$
\par
iii) $\Omega(x) = 0$  if  $x\in\mathcal{J}$; but $N_{\m}\equiv - \nabla_{\m}\Omega$ is nonsingular
in $\mathcal{J}$.
\par 
iv)Every null geodesic in $\mathcal{M}$ has two endpoints in $\mathcal{J}$.
\par
and, finally, the field equation:
\par
v)$R_{\m\n}\sim \lambda g_{\m\n}$
(cf. \cite{penrose}\cite{ashtekar}\cite{hawking}).
\par
(where we  have followed Penrose's notation $a\sim b$ to indicate two things that are
equal on $\mathcal{J}$ only).
\par
For example, for $AdS_4$ \cite{isham} one can choose coordinates in such a way that
\be
ds^2 =  sec^2 \rho ( d\tau^2 - d\rho^2 - sin^2 \rho( d\theta^2 + sin^2\theta d\phi^2))
\ee
where
\be
ds^2 =  d\tau^2 - d\rho^2 - sin^2 \rho( d\theta^2 + sin^2\theta d\phi^2)
\ee
is the metric of Einstein's static universe (ESU). We see that in this example
\be
\Omega\equiv cos \rho
\ee
and the {\em infinite} $\scri$ is located in these coordinates at the finite distance $\rho = \pi/2$.
\par
One of the simplest ways of characterizing the behavior of the conformal factor $\Omega$ is
to study null geodesics in its vicinity (cf. \cite{penrose}). We shall choose an affine parameter $\hat{u}$ on them,
(that is, $\hat{l}^{\m}\hat{\nabla}_{\m}\hat{ u} \sim 1$; where $\hat{l}^{\m}$ is the tangent (null) vector
to the geodesic)  and  we further fix the origin of the affine parameter by $\hat{u}\sim 0$.
\par
There is a corresponding parameter $u$  associated to the metric $g_{\m\n}$, defined by
\be
\frac{d\hat{u}}{d u} \equiv l^{\m}\nabla_{\m}\hat{u}\equiv \Omega^2 \hat{l}^{\m}\hat{\nabla}_{\m}\hat{u} = \Omega^2
\ee
\par
The conformal factor will have, by analiticity, an expansion of the type
\be
\Omega(\hat{u}) = -\sum_{n=1}^{\infty}A_n \hat{u}^n
\ee
If $\mathcal{M}$ is an Einstein space with scalar curvature given by
\be
R = \frac{2 n \lambda}{n - 2}
\ee
it is not difficult to show that the vector $N_{\m} \equiv - \nabla_{\m} \Omega$ obeys
\be
\hat{N}^2 \sim \frac{2\lambda}{(n - 1)( n - 2)}
\ee
conveying the fact that the sign of the cosmological constant is related to the spacetime properties of
the $\scri$ boundary (and, in particular, in the ordinary $(1,3)$ case, we see that
 $\mathcal{J}$ will be timelike when the cosmological constant is negative only).
\par
It is also possible to show that
\be
\hat{\nabla}_{\n}\hat{\nabla}_{\m}\Omega = \frac{1}{n} \hat{\Delta} \Omega ~\hat{g}_{\m\n}
\ee
which in turn, (using the fact that $l^2 = 0$), enforce $A_2 = 0$  in the preceding expansion (2.13)(\cite{penrose}).
\par
In order to make contact with the previous section, let us remark that we are here associating to a metric
in $\mathcal{M}$, a whole conformal class in $\mathcal{J}$.(That is, all the construction above is
invariant under $\Omega \rightarrow t~ \Omega$, with $t\in \mathbb{R}^{+}$).
\par
%
\par

\begin{figure}[!ht] 
\begin{center} 
\leavevmode 
\epsfxsize= 10cm
\epsffile{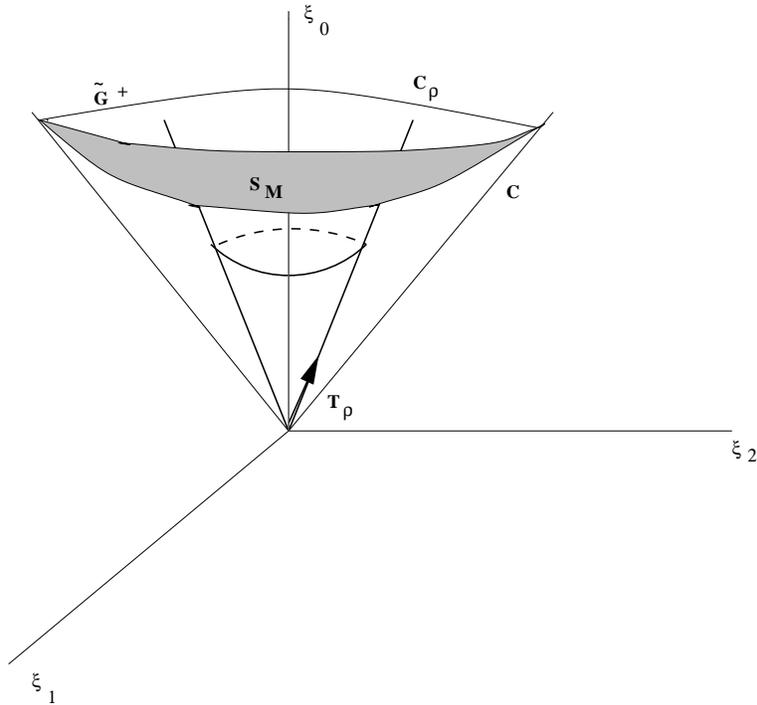} 
\caption{ $S_M$ in the simple case $M\equiv S^1$.}
\label{fig:dos} 
\end{center} 
\end{figure}

Let us now relate Penrose's conformal infinity construction with Lorentz holography as described in the previous subsection.
In order to fix the ideas let us come back to the example of Figure 1. We have indeed two complementary pieces
$\tilde{G}^{\pm}$ of $\tilde{G}$ with enjoy a common boundary, which we identify precisely with $G$.
The region $\tilde{G}^{+}$ will be identified with the interior of the light cone.  We can use  now the holographic coordinate
$\rho\in[-1,1]$ to parametrize the interior cones $\mathcal{C}_{\rho}$, as shown in the Figure~\ref{fig:dos}.
Let us denote by $\vec{T}_{\rho}$ the generator of dilatations at points located in the interior cone $\mathcal{C}_{\rho}$
It is clearly a vector along the cone $\mathcal{C}_{\rho}$. By identifying $\mathcal{C}$ with the light cone in $\tilde{G}$, we get
\be
\tilde{g}(\vec{T}_{\rho},\vec{T}_{\rho}) < 0
\ee
(for $\rho\in (0,1]$; that is, on $\tilde{G}^{+}$). This means that $\vec{T}_{\rho}$ is timelike in $\tilde{G}$. In order to
make contact with Penrose's construction let us mode out $\tilde{G}^{+}$ by the $\mathbb{R}^{+}$ action of the
dilatations. This can be simply done by imposing
\be
\tilde{g}(\vec{T}_{\rho},\vec{T}_{\rho}) = - 1
\ee
(again, when $\rho\in (0,1]$).
\par
In coordinates, it reads
\be
\xi_0^2 - \xi_1^2 - \xi_2^2 = 1
\ee
This defines a surface in $\tilde{G}^{+}$, which will be denoted by $S_M$. This is a manifold whose
 boundary in Penrose's sense is identical to the manifold $M$. In the particular case $M\equiv S^1$, the manifold $S_M$
corresponds to the dashed region in the Figure~\ref{fig:dos}, a region that brings to mind at once the AdS example.
\par
The metric induced in $S_M$ by its imbedding in $\tilde{G}$ can be easily computed by a slight generalization
of the usual horospheric coordinates. Actually, for arbitrary $\pm$ signs, denoted by $\epsilon_i = \pm 1$, the metric induced on the surface
\be
\sum_{i = 1}^n  \epsilon_i x_i^2 = 1
\ee
by the imbedding on the flat space with metric
\be
ds^2 =\sum_{i = 1}^n  \epsilon_i d x_i^2 
\ee
can easily be reduced to a generalization of Poincar\'e's metric for the half-plane by introducing the coordinates
\bea
&&z\equiv x^{-}\nonumber\\
&&y^{\m}\equiv z ~x^{\m}
\eea
where we have chosen the two last coordinates, $x^{n-1}$  and $ x^n$ in such a way that their contribution to the metric
is $ dx_{n-1}^2 - dx_n^2$ (this is always possible if we have at least one timelike coordinate); and we define
$x^{-}\equiv x^{n} - x^{n-1}$. $\m\in (1,\ldots n-2)$.
The generalization of the Poincar\'e metric is:
\be
ds^2 = \frac{\sum \epsilon_{\m} dy_{\m}^2 - dz^2}{z^2}
\ee
It should be clear by the  reasoning above that when $M$ is of signature $(p,q)$, $S_M$ enjoys signature $(p,q+1)$.
In local coordinates $(x,t,\rho)$, the metric $\tilde{g}$ on $\tilde{G}^{+}$ can be written as
\be
d\tilde{s}^2 = t^2 ds^2(x,\rho) - dt^2
\ee
implying at once (whenever $n>1$)
\be
\tilde{R}_{\m\n} = R_{\m\n} + (n-1)  g_{\m\n}
\ee
This means that when $\tilde{G}^{+}$ is Ricci flat, $S_M\sim \tilde{G}^{+}/\mathbb{R}^{+}$
is an Einstein space with  negative cosmological cosntant.
\par
We could, {\em mutatis mutandis} repeat the previous construction for $\tilde{G}^{-}$. We would impose the condition
\be
\tilde{g}(\vec{T}_{\rho},\vec{T}_{\rho}) = 1
\ee
($\rho\in [-1,0)$) The Figure~\ref{fig:dos} clearly shows that the main difference with the previous construction is that
now we are forced to assume that $M\equiv S^1$ has signature $(1,0)$ (instead of $(0,1)$ as we had implicitly assumed
before). Furthermore, the line element would contain $+dt^2$ (instead of $- dt^2$). All this implies that $S_M$ is now
 an Einstein
space with positive cosmological constant. 
\par
In general this would mean that the holographic coordinate is in this case timelike, and therefore
in order to get a boundary with the desired signature $(p,q)$ we should consider a bulk space of signature $(p+1,q)$ (instead
of $(p,q+1)$).
\par
The preceding remark is potentially interesting for a {\em boundary} physical spacetime of Minkowskian
 signature $(1,3)$, since in this case we could try to perform a holographic projection with positive cosmological constant,
but on a bulk spacetime with signature $(2,3)$.
\par
Let us now summarize the   {\tt rules of the geometric holography}.
\begin{itemize}
\item Given the manifold $M$ of signature $(p,q)$, we first build a Lorentzian
 ambient space $\tilde{G}$ of signature $(p+1,q+1)$
\item In the ambient space $\tilde{G}$ look for a {\em light cone} defined relative to the extra time in $\tilde{G}$ and
with transversal sections isomorphic to the manifold $M$ we started with. This light cone is naturally associated with
a conformal class in $M$.
\item The {\em bulk} space-time is now a manifold $S_M$ of signature $(p,q+1)$ inside the light cone and with boundary 
a transversal section which should be identified with $M$. This bulk space is simply defined as the quotient of the space 
interior to the light cone by dilatations on the ambient spacetime $\tilde{G}$.
\end{itemize}
Actually, the previous set of holographic rules do allow in principle a further generalization of the holographic map to 
non-local variables.

\subsection{Non-Local Holography: Wilson Loops}.

A first step in order to extend holography to non-local loop variables on $M$ would consist in associating to a closed loop
$C$ located in $M$, a two-dimensional manifold, $S_C$, with boundary $C$ itself, and such that $S_C$ is contained
in the region $S_M$ precisely defined in the previous paragraph.
\par
A natural definition of $S_C$ would be the locus of points previously defined in eq. (1.20), but now reducing
 to vectors $\vec{T}_{\rho}$
associated with dilatations at points $(x,\rho)$ such that $x\in C$. It is plain that the manifold $S_C$ would be
precisely of the type depicted in the Figure \ref{fig:dos} in the simplest example $M\equiv S^1$. It is this holographic 
manifold $S_C$ that from a physical point of view, we would like to associated to a string world sheet. 
\par
Actually, the preceding construction already conveys the conformal invariance of the holographic representation
of the Wilson loop. If we define the vacuum expectation value of the Wilson loop the area of the surface $S_C$
measured in the $\tilde{g}$ metric, it is clear that it will only depend on the conformal class of the metric on $M$ and therefore
it would be, by construction, independent of conformal transformations of the loop $C$. (This is actually well known for
 Wilson loops in the AdS/CFT case)\cite{wilson}.
\par
From this point of view, the fundamental zig-zag invariance \cite{zz}( which physically means that the
 vacuum expectation value
 of the loop is not sensitive to foldings of the loop),  that is,  invariance under orientation-reversal
diffeomorphisms of the loop itself (located at  $M$), is perhaps related to some bigger invariance present
already in the bulk.
We hope to come back to this issue in the future.


\section{Bondi News and the Holographic Map}

The geometric holography just described strongly depends on the existence of a unique solution for a Cauchy problem
defined in terms of Einstein's equations on the bulk and with initial conditions fixed by the conformal class of the physical
spacetime metric at the boundary. A necessary condition for the uniqueness of the solution is the vanishing of
(gravitational) Bondi news through $\mathcal{J}$. The physical meaning of this is that the spacetime boundary
$\scri$ is {\em opaque} to gravitational radiation; in the four-dimensional case, with topology $S^3\times \mathbb{R}$
absence of Bondi news requires that the Bach tensor vanishes on $\scri$. Recall that this construct is defined
 (in any dimension) using the tensor
$A_{\a\b}\equiv\frac{1}{n-2}(R_{\a\b} - \frac{R}{2(n-1)} g_{\a\b})$  as:
\be
B_{\m\n}\equiv \nabla^{\rho}C_{\rho\m\n} + A^{\a\b}W_{\a\b\m\n}
\ee
where $W_{\a\b\m\n}$ is the Weyl tensor, defined in terms of the Riemann tensor as:
 $W_{\a\b\m\n}\equiv R_{\a\b\m\n}- (A_{\b\m} g_{\a\n} + A_{\a\n}g_{\b\m}  - A_{\b\n}g_{\a\m}  - A_{\a\m}g_{\b\n})$
(so that it vanishes when n=2 or n=3).

and $C_{\m\n\rho}$ is the Cotton tensor, defined by:
\be
C_{\a\b\g}\equiv \nabla_{\a}A_{\b\g} - \nabla_{\b} A_{\a\g}
\ee
In n=3 the spacetime is conformally flat iff the Cotton tensor vanishes. It is to be stressed that the Bach tensor is conformally
invariant  when n=4 {\em only}.
\par
When n=3 (with topology $S^2\times \RR$), the Bach tensor vanishes, conveying the fact that there are no gravitational Bondi
news in this case (\cite{ashtekar}). This is the simplest instance of a general theorem proved in \cite{fg} staying that for a 
spacetime $M$ of even dimension  there is no obstruction for the existence of a formal power series solution to the
Cauchy problem.
\par
In the case of a n=3 boundary spacetime $S^2\times \RR$, however, Bondi news exist in general for matter fields if the Cotton tensor
does not vanish. This means that for four dimensional bulk spaces there is the possibility of having a well defined Cauchy problem
in the Fefferman-Graham sense, and yet, Bondi news for fields with spin different from 2. It is obvious that this is problematic from the
holographic point of view (except in the case of pure gravity).
\par
Geometrically,  the vanishing of the Cotton tensor in the three-dimensional case is the necessary and
 sufficient condition for the existence of conformal Killing spinors. (In the four dimensional case, the equivalent condition 
(implying that the space is conformally Einstein)  is
 the vanishing of the Bach tensor (cf.\cite{kozameh})). Only in this case the conformal symmetry is realized asymptotically
in such a way that one can define asymptotically conserved charges associated to the $O(3,2)$ conformal group. ($O(4,2)$ in
the four-dimensional case).
\par 
This reduction of the asymptotic symmetry group to AdS is similar to the reduction from the asymptotic Bondi-Metzner-Sachs
\cite{penrose} group in the asymptotically flat case, towards Poincar\'e, as has been pointed out in \cite{ashtekar}. In this case, however,
the condition $B_{\m\n}\sim 0$ is too strong, and, in particular, it is not stable against gravitational perturbations.
There is then a curious discontinuity in the limit $\lambda\rightarrow 0$.
\par
A curious fact is that it would seem that the vanishing of the conformal anomaly in the three dimensional case
in the holographic setting \cite{skenderis} does not need the vanishing of the Cotton tensor.
\par
Another fact worth stressing is that gravitational Bondi news are generically non-vanishing when $\scri$ is spacelike,
or even null. It is plain that the interplay between holography and Bondi news is related to the existence of a Cauchy surface 
for asimptotically anti de Sitter spacetimes (i.e. $\scri$ timelike).
The simplest example is, obviously, AdS itself, where in order to define a Cauchy surface one is forced to impose 
{\em reflective} boundary conditions on $\scri$,\cite{isham}
 enforcing the desired absence of Bondi news for matter fields..
\par
Remarkably enough, Hawking \cite{hawking} has proved that the physics of this  set of boundary conditions
is equivalent to assuming that the gravitational fields tend to AdS at infinity fast enough.
Physically, absence of Bondi news on $\scri$ is necessary in order that a CFT living on $\scri$
could {\em propagate} holographically to the bulk in a unique way.

\section{Renormalization Group Flow and Einstein Equations}

One of the deepest results of the world sheet sigma model approach to string
theory is the interpretation of the background spacetime metric as a coupling constant
and Einstein's equations in spacetime as the corresponding renormalization group beta functions
(\cite{polyakov}). The extension to this approach to the physical spacetime metric for ordinary
four dimensional quantum field theory is maybe possible provided a holographic projection
works.
\par
In a holographic framework one considers generically an ambient bulk spacetime where the real
{\em physical} spacetime is somehow embedded as a subspece of codimension 1. We shall denote
generically by $\rho$ this holographic coordinate.
\par
Let us now consider the one-dimensional family of induced four-dimensional metrics, $g_{\m\n}(\rho)$
which can be considered as well as coupling constants. This leads to an interpretation
of the $\rho$-dependence as a {\em flow} in the Renormalization Group sense.
\par
From the geometric point of view described in previous sections, holography reduces essentially to 
determine this $\rho$ dependence of the four-dimensional metric in terms of the Einstein
 equations for the bulk geometry. From a physical point of view, one of the most fascinating aspects
 of the whole set up of holography is the re-interpretation of the renormalization group equations of
 the quantum field theory living at the boundary in terms of Einstein's equations in the bulk. Following
Maldanena's \cite{malda} work on near-horizon geometry of D-branes the holographic coordinate
$\rho$ can be related to the ultraviolet scale of the theory.
\par
We would now like to be more specific about the physical meaning
of the $\rho$ dependence of the four-dimensional metric. It will acually be proposed to use an appropiate
 function of $g_{\m\n}(\rho)$
as a way of defining Zamolodchikov's c-function \cite{zamo}  of the theory at the {\em scale }$\rho$
(more precisely, as a way to measure the relevant massless degrees of freedom at the scale
$\rho$). Based on this interpretation, generic geometric properties of the bulk geometry
and the holographic equations suffice to ensure the validity of the c-theorem, to wit, that
the number of massless degrees of freedom decreases through the renormalization group flow
from the ultraviolet (UV) towards the infrared (IR).
An importat point will be to understand under what conditions holography is consistent with departing from the 
CFT fixed point.
\par
In order to address this last question it is important to keep in mind the two different variables
used in the Lorentzian (namely, $\tilde{G}$ as bulk space) version of the holographic geometry.
One variable $t\in \mathbb{R}^{+}$ characterizes conformal transformations of the spacetime metric,
while the variable $\rho\in [-1,1]$ (that is, the one we can properly interpret as the holographic coordinate)
parametrizes the change of the metric in the bulk direction.
\par
The interplay between these two variables can be easily undestood in Penrose's framework.
In fact we have already seen that close to $\scri$ we have

\be
\Omega(\hat{u}) = -\sum_{n=1}^{\infty}A_n \hat{u}^n
\ee
 so that if all $A_n = 0$ for $n > 1$, rescalings of the holographic variable $\hat{u}\rightarrow \lambda\hat{u}$
are equivalent to conformal transformations of the metric. This shall be the typical situation of what we can call
{\em conformal holography}, or in other words holography describing CFT fixed points.
\par
In the general case, when no coefficient vanish in the above expansion, this simple equivalence
does not hold anymore and one can have a nontrivial renormalization group flow. Very likely the difference between
the conformal and the non-conformal case is related to the presence of an energy momentum tensor in the 
second member of the bulk
Einstein's equations.
\par
After these preliminary remarks on the holographic interpretation of the renormalization group flow, let us proceed
to discuss in detail the holographic definition of the c-function.


\section{A Proposal for a Holographic c-Function}

In \cite{zamo} Zamolodchikov proved that for local and unitary two-dimensional field theories
there exists a function of the couplings , hereinafter  called the c-function, $c(g_i)$, such that
\be
- \b_i \pd^i c \leqslant 0
\ee
along the renormalization group flow. For fixed points of the flow, the c-function reduces to the central extension
of the Virasoro algebra. Some generalizations of the c-function to realistic four-dimensional theories
have been suggested; let us mention in particular Cardy's proposal on $S^4$:
\be
c \equiv \int_{S^4} \sqrt{g} < T >
\ee
where $T$ is the trace of the energy-momentum tensor of the theory. There is no complete agreement as to whether
a convincing proof of the theorem exists for dimension higher than two (cf. \cite{latorre} for a recent attempt).
\par
It is not difficult to invent a holographic definition of the central extension for N=4 super Yang-Mills
theories using the AdS/CFT map \cite{gkp}\cite{map}, especially in the light of the IR/UV connection
pointed out by Susskind and Witten in \cite{sw}.
In order to understand this construction properly, let us recall the original 't Hooft's presentation of the holographic
principle \cite{'th}, stemmning from the Bekenstein-Hawking entropy formula for black holes \cite{bh}.
Given a bounded region of instantaneous $d-1$  space $V$  of volume $vol_{d-1}(V)$, holography states
 that all the physical information on processes in $V_{d-1}$
can be codified in terms of surface variables, living on the boundary of $V$, $\pd V$. 
More precisely, the number of holographic degrees of freedom is given by:
\be
N_{d.o.f.} \sim \frac{vol_{(d-2)}(\pd V)}{G^{(d)}}
\ee
Physically this means that we have precisely one degree of freedom in each area cell of size given by the Plack length.
In spite of the fact that the Bekenstein bounds would suggest the radical approach that {\em any} physics
in $V$ can be mapped into holographic degrees of freedom in $\pd V$, the list of theories suspected to admit
holographic projection is still small, and always involves gravity. (Susskind indeed suggested  from the beginning 
that string theory
should be holographic).
\par
In the previous sections of the present paper we have just examined the geometrical set up in which to a given geometrical
(diff-invariant) theory defined in the bulk one can associate a CFT living on the boundary. Under these conditions, it is not
difficult to generalize 't Hooft's formula in a precise manner.

\par
Let us consider a four dimensional CFT defined on a spacetime with topology $S^3\times\mathbb{R}$ and with the natural metric
in $S^3$. Let us also introduce an ultraviolet cutoff $\d$, and let us correspondingly divide the sphere $S^3$ into
small cells of size $\d^3$. The number of cells is clearly of order $\frac{1}{\d^3}$. We would like now to define the number of degrees of
freedom in terms of the central extension as
\be
N_{dof} = \frac{c}{\d^3}
\ee
 Notice that here the parameter $c$ plays the r\^ole of the number of degrees of freedom in each cell. If the theory we are considering
is the holographic projection of some supergravity in the bulk, it is natural to rewrite this in terms of (5.23), but with the $vol(\pd V)$ 
now replaced by a section of the bulk at $\d = constant$, with $\d$ being now identified with the holographic
parameter. We are thus led to the identification
\be
c\equiv lim_{\d\rightarrow 0} \frac {\d^3 Vol(\pd V_{\d})}{G_5}
\ee
In the particular example of AdS this yields $c = \frac{R^3}{G_5}$, with $R^4 = \alpha'^{2} N g^2$\cite{malda}.

\section{Renormalization Group Flow along Null Geodesics}

We shall in this section study the renormalization group  evolution of the postulated c-function ;
that is, its dependence on  the holographic variable, $\rho$.
In order to do that, the first point is to identify exactly what we understand by {\em area} (that is,
$vol(\pd V)$. Our definition clearly involves the
quotient between an area defined close to the horizon and an {\em inertial area}, so that:
\be
c(\d)\equiv  \frac{vol_{(d-2)}(\scri_{\delta}) \times vol_{(d-2)}(inertial)}{G_d}
\ee
The meaning of the preceding formula is as follows (cf. \cite{sw}). The first term, $vol_{(d-2)}(\scri_{\delta})$ is the volume
computed on $\scri$ regularized with an UV cutoff $\delta$ ($N^2/\delta^3$ in the familiar example of $ AdS_5$).
The other factor, $vol_{(d-2)}(inertial)$ stands for the equivalent volume measured by an inertial observer which does not
feel the gravitational field (that is, $\delta^3$ in $AdS_5$). The whole thing is then divided by the d-dimensional Newton's constant.
\par
Our c-function will obey a renormalization group equation of the type:
\be
\d\frac{\pd c}{\pd\d} + \sum_i \b_i(g)\frac{\pd c}{\pd g_i} = 0
\ee
which means that to prove the c-theorem we just have to show that:
\be
\d\frac{\pd c}{\pd\d}\leq 0
\ee
\par
Now, to study the evolution in the bulk of the c-function, we need to know how this regularized definition of area evolves as we
 penetrate into the bulk. It is now only natural to identify the UV cutoff $\delta$ with the affine parameter $\hat{u}$
introduced in the previous section such that $\hat{u}\sim 0$.
\par
We would like also to argue that it is quite convenient to study the evolution of the area along null geodesics entering the bulk.
First of all,the whole set-up is conformally invariant (which is not the case for timelike geodesics).
In addition, there is a very natural definition of tranverse space there. In a Newman-Penrose orthonormal tetrad 
\footnote{We choose to present the formulas in the four dimensional case by simplicity, but it should be clear that no 
essential aspect depends on this.},
which
is a sort of complexified light cone, because in terms of a real orthonormal tetrad, $e^a$,
\bea
l^{\m}\pd_{\m}&\equiv & e^{+} \equiv \frac{1}{\sqrt{2}}(e^0 + e^3)\nonumber\\
n^{\m}\pd_{\m}&\equiv & e^{-} \equiv \frac{1}{\sqrt{2}}(e^0 -e^3)\nonumber\\
m^{\m}\pd_{\m}&\equiv  & e_{T}\equiv\frac{1}{\sqrt{2}}(e^1 - i  e^2)\nonumber\\
\bar{m}^{\m}\pd_{\m}&\equiv &\bar{e}_{T}\equiv \frac{1}{\sqrt{2}}(e^1 + i e^2),
\eea
one can easily find the optical scalars \cite{kramer} of the geodesic congruence. One has, in particular, that
\be
\rho \equiv - \nabla_{\m} l_{\n} m^{\n}\bar{m}^{\m} = -(\theta + i \omega)
\ee

where the {\em expansion}, $\theta$, is defined by $\theta\equiv \frac{1}{2}\nabla_{\a} l^{\a}$ and the {\em rotation},
$\omega$,   is a scalar which measures the 
antisymmetric part of the covariant derivative of the tangent field: $\omega^2\equiv \frac{1}{2}\omega_{\a\b}\omega^{\a\b}$, with
$\omega_{\a\b}\equiv\nabla_{[\a}l_{\b]}$. 
\par
Let us now consider a  {\em congruence } of null geodesics. This means that we have a family $x^{\m}(u,v)$, such that
$v$ tells in which geodesic we are, and $u$ is an affine parameter of the type previously considered. The connecting vector
({\em geodesic deviation})
$Z^{\m}\equiv x^{\m}(u,v) -  x^{\m}(u, v + \delta v) $ connects points on neighboring geodesics, and by construction
satisfies
\be
\pounds (l)Z^{\m} = 0
\ee
that is, $l^{\m}\nabla_{\m}Z^{\a} = Z^{\m}\nabla_{\m}l^{\a}$. Although the molulus of the vector $Z$ is itself not conserved, 
it is not difficult to show that its projection on $l^{\m}$ is a  constant of motion. Penrose and Rindler call {\em abreast} the congruences for 
which this projection vanishes. In this case one can show that $h = 0$, where h is defined from the projection of the
geodesic deviation vector on the Newman-Penrose tetrad:
\be
Z^{\a} = g~ l^{\a} + \zeta ~\bar{m}^{\a} + \bar{\zeta}~m^{\a} + h~ n^{\a}
\ee
Under the preceding circumstances, the triangle $(0,\zeta_1,\zeta_2)$ is contained in $\Pi$, the 2-plane 
spanned by the real and imaginary parts of $m^{\a}$.
\footnote{In the general case, it is plain that in this way we build a $d-2$-volume}. Now it can be proven \cite{penrose} that, calling
$A_2$ the area of this elementary triangle,
\be
l^{\a}\nabla_{\a} A_2 = -(\rho + \bar{\rho}) A_2  =  2 \theta~ A_2
\ee
This fact relates in a natural way areas with null geodesic congruences.
\par
Using this information we can write at once:
\be
\hat{u}\frac{d c(\hat{u})}{d\hat{u}} = (\theta \hat{u}+ d-2 )c(\hat{u})
\ee
It is worth noting at this point that $\hat{\theta}$ is finite (it corresponds to Einstein's static universe in the
standard AdS example). The divergence in $\theta$ stems from the conformal transformation necessary to go from $\hat{g}_{\a\b}$
to $g_{\a\b}$, to wit:
\be
\theta = \hat{\theta} + \frac{d-2}{2} \frac{N.Z}{\Omega}
\ee
 In order that inertial and $\scri$ units be the same, it is natural to measure inertial areas in units of $\delta \equiv \frac{\hat{u}}{
(N.Z)_{\scri}}$, where $ (N.Z)_{\scri}$ represents the scalar product of the vector $N^{\m}\equiv - \nabla^{\m} \Omega$ and
$Z^{\m}$ computed at $\hat{u} = 0$.
Doing that one gets that the first derivative  vanishes to first order:
\be
\hat{u}\frac{d c(\hat{u})}{d\hat{u}} = 0
\ee
  But we can now invocate a well known theorem by Raychadhuri \cite{r}\cite{hellis}
\be
l^{\m}\nabla_{\m} \theta =  \omega^2 - \frac{1}{2}R_{\m\n}l^{\m}l^{\n} -  \sigma \bar{\sigma} - \theta^2
\ee
(where  the {\em shear} $\sigma\bar{\sigma} \equiv \frac{1}{2}\nabla_{[\b}l_{\a]} \nabla^{[\b}l^{\a]}
 - \frac{1}{4}(\nabla_{\a}l^{\a})^2)$
\par
The  Ricci term in the above equation vanishes for Einstein spaces, and the rotation must necessarily be zero
if we want the flow lines to be orthogonal to the surfaces of transitivity; that is, that there exists a family of hypersurfaces
$\Sigma$, such that $l_{\m} = \nabla_{\m}\Sigma$.
\par
This shows that under these conditions
\be
\hat{u}\frac{d\theta}{d\hat{u}} < 0
\ee
which is enough to prove the c-theorem in the holographic case.

\section{Final Comments}
We have considered only the null infinity $\scri$  in this paper. All particles ending there
are massless. It would be exceedingly interesting
to extend the geometrical holographic analysis to the massive case. This would presumably mean, in Penrose's language, to consider
the {\em timelike infinity}, $i^{+}$. This would answer, in particular, a natural question the reader might like to ask, and that is
why we do not see the sphere $S^5$ in our approach (in the simplest case of $AdS_5$).
 We think that we would see the sphere in  the sense of seeing the higher modes of the  allowed 
supergravity Kaluza-Klein spectrum
on AdS through boundary conditions at timelike infinity.
\par
Zig-zag symmetry, believed to be an exact symmetry of Wilson loops, does not seem to be naturally implemented in the geometrical approach.
It appears that some generalization of it is needed, which is in some sense purely topological at the boundary.
\par
We have mentioned in several places that in some cases bulk spaces could exist with {\em positive} cosmological constant. This would mean,
in particular, that the Minkowskian holography would correspond to a $(2,3)$ signature five-dimensional bulk manifold.
An explicit construction would be most welcome.
\par
Work is in progress in these, and related, matters.

\section*{Acknowledgments}
We are grateful for stimulating discussions with L. Alvarez-Gaum\'e, G. Gibbons. M. Henningson, J.I. Latorre,
H. Osborn and E. Rabinovici.
This work ~~has been partially supported by the
European Union TMR program FMRX-CT96-0012 {\sl Integrability,
  Non-perturbative Effects, and Symmetry in Quantum Field Theory} and
by the Spanish grant AEN96-1655.  The work of E.A.~has also been
supported by the European Union TMR program ERBFMRX-CT96-0090 {\sl Beyond the Standard model} 
 and  the Spanish grant  AEN96-1664.


\appendix


\end{document}